\documentclass[aps,twocolumn,superscriptaddress,prb]{revtex4}
\usepackage{amsmath}
\usepackage{amssymb}
\usepackage{enumerate}

\newcommand\av[1]{\left\langle {#1}\right\rangle }
\newcommand\gN[1]{\langle \!\langle g^{#1}\rangle \!\rangle }
\newcommand\cum[1]{\langle \!\langle {#1}\rangle \!\rangle }

\newcommand\gB[1]{\langle {g^{#1}}\rangle{\lower
4pt\hbox{$ \scriptstyle{\!B} $}}}
\newcommand\avB[1]{\langle {#1}\rangle {\lower
4pt\hbox{$ \scriptstyle{\!B} $}}}
\newcommand\gBc[1]{\langle \!\langle {g^{#1}}\rangle\!\rangle {\lower
4pt\hbox{$ \scriptstyle{\!B} $}}}

\newcommand\dgB[1]{\langle \!\langle {{\delta}g^{#1}}\rangle\!\rangle {\lower
4pt\hbox{$ \scriptstyle{\!B} $}}}

\newcommand\cumB[1]{\langle \!\langle {{#1}}\rangle\!\rangle {\lower
4pt\hbox{$ \scriptstyle{\!B} $}}}

\newcommand\dd{{\operatorname{d}}}
\newcommand\e{{\operatorname{e}}}
\newcommand\Ref[1]{Ref.\ [\onlinecite{#1}]}
\newcommand\Refs[1]{Refs.\ [\onlinecite{#1}]}
\newcommand\Eq[1]{{Eq.~(\ref{#1})}}
\newcommand\EQS[2]{{Eqs.\ (\ref{#1}) and (\ref{#2})}}
\newcommand\lr[3]{\left#1{#3}\right#2}
\newcommand\Lr[4]{\left#1{#3}\right#2^{\!#4}}

\begin{document}

\title{On the Applicability of the Ergodicity Hypothesis  to
Mesoscopic Fluctuations}
\author{O. Tsyplyatyev}
\affiliation{Department of Physics, Lancaster University,
Lancaster LA1 4YB, United Kingdom}
\author{I.~L.\ Aleiner}
\affiliation{Physics Department, Columbia University, New York, NY
10027}
 \author{Vladimir I. Fal'ko}
\affiliation{Department of Physics, Lancaster University,
Lancaster LA1 4YB, United Kingdom}
 \author{Igor V. Lerner }
 \affiliation{School of Physics and Astronomy, University of
Birmingham,
 Birmingham B15 2TT, United Kingdom}
\date{\today}

\begin{abstract}
We evaluate a typical value of higher order cumulants (irreducible
moments) of conductance fluctuations that could be extracted from
magneto-conductance measurements in a single sample when an
external magnetic field is swept over an interval $B_0$. We find
that the n-th cumulant has a sample-dependent random part $\pm \gN
2^{n/2}\sqrt{a_{n}B_{c}/B_{0}}$, where $\gN 2$ is the variance of
conductance fluctuations, $B_{c}$ is a correlation field, and
$a_{n}\sim n!$. This means that an apparent deviation of the
conductance distribution from a Gaussian shape, manifested by
non-vanishing higher cumulants, can be a spurious result of
correlations of conductances at different values of the magnetic
field.
\end{abstract}

\maketitle

The ergodicity hypothesis (EH) plays a crucial role in mesoscopic
physics. It identifies mesoscopic `sample-to-sample' fluctuations
of a certain quantity with an apparently random but reproducible
spread in the magnitude of this quantity measured in a single
sample as a function of some controllable parameter, e.g., an
external magnetic field or a gate voltage. The EH is based on the
idea that, since the fluctuations result from the electron
interference in scattering from impurities,
\cite{Al:85,AlKh:85,L+S:85} changing the interference conditions
by, e.g., sweeping the field over a wide range, one obtains a data
set (usually called a magnetofingerprint) statistically equivalent
to that for sample-to-sample fluctuations.

Since most of experiments are performed, by necessity, on a few
samples  while practically any theoretical approach  uses some
form of the ensemble averaging over disorder,\cite{ALW} the EH
provides the framework for comparing experiment with theory in
mesoscopics. Such a framework appears to work so well that it is
now routinely used by default, in particular in experiments
\cite{Chang:96a,Marcus:96a,MW,Swastik} aiming to determine the
shape of the conductance distribution.

Here we  show that for the ergodicity hypothesis to be valid for
analyzing an apparent deviation of the distribution from the
normal shape in systems with large conductance,  one needs the
range of magnetic fields so large that it may not be reachable in
modern experiments. We show that a non-Gaussian shape of the
distribution extracted from a single-sample magnetofingerprint can
be an entirely spurious result of the non-ergodicity  due to
residual correlations of conductances at different values of the
magnetic field. For brevity, we consider only conductance
distributions, although our method and results can be extended to
any mesoscopic observable.

Let us start with a brief description of theoretical predictions
on the conductance distribution. Any distribution can be described
by its moments or cumulants (irreducible moments). \cite{AdvStat}
Within the standard one-parameter scaling approach to disordered
conductors, \cite{AALR} the cumulants of conductance fluctuations
in the metallic regime at temperature $T=0$ (i.e.\ for a fully
phase-coherent conductor) are given
by\cite{AKL:86,MPS,comment1}
\begin{equation}\label{gn}
    \gN n=A_{nd}\av g^{2-n}\,,
\end{equation}
where $A_{nd}$ are numerical coefficients dependent on the
cumulant order $n$ and dimensionality $d$. Here $\av g$ is the
ensemble-averaged conductance (i.e.\ conductance averaged over all
the realizations of disorder) measured in the units of $e^{2}/h$.
In a good metal $\av g\gg1$, and the higher ($n>2$) cumulants in
\Eq{gn} are small\cite{comment2} in comparison with the universal
variance (the second cumulant), $\gN2\equiv \av{ {g}^{2}}-\av
g^2=A_{2d}\sim1$. A characteristic feature of the Gaussian
(normal) distribution is that all its high cumulants vanish.
Therefore, \Eq{gn} means that the distribution of conductance
fluctuations is expected to be almost Gaussian. An experimental
detection of deviations from the Gaussian shape, characterized by
the higher cumulants in \Eq{gn} represents a challenging task.

This is even more so, when one realizes that at $T\ne0$ \Eq{gn} is
only valid for sample sizes smaller that phase-breaking length
$L_\phi$. Thus for a quasi-1D wire with length $L>L_\phi$,
conductance fluctuations can be estimated as resulting from
independent contributions of $N=L/L_\phi$ conductors connected in
series, which lead, in accordance with the central limiting
theorem,\cite{AdvStat} to a further suppression of higher
cumulants:\cite{comment3}
\begin{equation}\label{Lphi}
    \gN n_{_{L>L_\phi}}\sim\gN
    n_{_{L=L_\phi}}\Lr(){\frac{L_\phi}{L}}{2n-1}\,.
\end{equation}
However small are the deviations from the Gaussian statistics
described by \EQS{gn}{Lphi}, they would be in principle
experimentally detectable, had one been able to accumulate
sufficient statistical data from uncorrelated different
realizations of disorder -- a task which is in principle possible
for semiconductor nanostructures by applying a large number of
annealing cycles.\cite{Sanquer} Our main point is that the failure
of the EH invalidates an apparently simpler alternative (and the
only one known for metallic wires) based on extracting statistical
data from a magnetofingerprint $g(B)$.

To explain how this occurs, let us formalize a procedure of
extracting the moments $\gB n$ of conductance distribution from
$g(B)$. It is done by sampling it into a histogram, so that $\gB
n$ are determined by the following averages\cite{comment4} (where
a smoothed histogram $f_B(g)$ is the distribution function):
\begin{equation}\label{gBn}
\gB n\equiv \int_{{\tilde B}}^{{\tilde B}+B_{0}}g^n(B)\frac{\dd
B}{B_{0}}\equiv \int g^n (B) f_B(g)\frac{\dd B}{B_{0}}\,.
\end{equation} The
cumulants $\gBc n$ are then found from these (reducible) moments
using the standard relations\cite{AdvStat}, see \Eq{irreduc}
below.

It is well known\cite{AlKh:85,L+S:85}
that for $B\,,B'\gg B_c$ values of $g(B)$ are correlated for
different $B$ with the co-variance
\begin{equation}
\kappa (B-B^{\prime })\equiv \cum{g(B)g(B^{\prime })} =\gN2 \times
K\!\lr(){\frac{B-B^{\prime }}{B_{c}}},\; \label{Covar}
\end{equation}%
where  $K(x)$ decays slowly (as a power-law, see \Eq K below) from
$K(0)=1$  to $0$. Its form and the value of the correlation field
$B_{c}$ (which corresponds to one magnetic flux quantum through
the region of coherence) depend on the dimensionality, temperature
regime, size and geometry of the system. The ergodicity has
formally been proved\cite{AKL:86a} in the limit
$B_0/B_c\to\infty$. For a finite $B_0$, one needs to evaluate to
what extent the correlations (\ref{Covar}) contribute to $\gBc n$.
To estimate a typical value of $\gBc n$ in a given sample, we
calculate its disorder-averaged mean square, $\av {\gBc n^2}$.

 In the leading order in
$B_{c}/B_{0}\ll 1$, we find
\begin{equation}
\label{result}\begin{aligned} \av {\gBc n^2} &=\gN
n^2+a_n\frac{B_{c}}{B_{0}}\,\gN2 ^{n}\!\\
a_n&=n!\int_{-\infty }^{\infty }\left[ K(x)\right] ^{n}\dd x
\end{aligned}%
\end{equation}
This equation is the main result of the paper. It shows that the
ergodic hypothesis works only for the variance itself, as $\av
{\gBc 2^2} =A\gN2^2$ with the coefficient of proportionality $A$
being very close to $1$ for $B_{c}/B_{0}\ll 1$. For $n\ge3$, the
second term in \Eq{result} represents a systematic error in
extracting a true value of the cumulant from the
magnetofingerprint. No sampling procedure in the statistical
analysis of a magnetofingerprint would produce error bars smaller
than this systematic error. In spite of the small parameter
$B_{c}/B_{0}$, this error term is dominant for measurements in
highly conducting wires due to a very fast decrease of the
cumulants with $n$, \Eq{gn}, further augmented for samples with
$L>L_\phi$, \Eq{Lphi}.

Thus even a perfectly Gaussian sample-so-sample distribution can
generate a visibly non-Gaussian one extracted from the
magnetofingerprint of a single sample, owing to the contribution
of conductance correlations in \Eq{result}. Experimental values of
cumulants extracted in this way are sound only if they comfortably
exceed the systematic error,
$\pm \gN2 ^{n/2}\,\sqrt{%
a_{n}B_{c}/B_{0}}$. This is manifestly not the case for
experimental results of \Ref{MW}. On the other hand, in
multi-parametric conductance fluctuations experiments, such as in
\Refs{Chang:96a,Marcus:96a}, where realizations are effectively
changed also by a sweeping the gate voltage (and thus changing the
Fermi wavelength) over some interval $V_0$, such a systematic
error may be reduced down to $\pm \gN2 ^{n/2}\,\sqrt{\left(
B_{c}/B_{0}\right) \left( V_{c}/V_{0}\right) }$, where $V_c$ is a
correlation value of the gate voltage.

Before deriving \Eq{result}, let us recall some basic relations
between moments and cumulants:\cite{AdvStat}
\begin{equation}
\frac{\av{ g^{n}}}{n!} =\sum_{\{m_{i}\}}\left\{
\prod_{k=1}^n\frac{1}{m_{k}!}\left( \frac{\gN k}{k!}\right)
^{\!m_k}\right\} . \label{irreduc}
\end{equation}%
The summation here is carried out over all the partitions
$\lr\{\}{m_i}=\lr\{\}{m_1,\ldots,m_n}$ of $n$ into the sum $n=
\sum_{k=1}^nkm_k$, with $m_k$ being non-negative integers. Note
that in a  diagrammatic approach\cite{AKL:86} cumulants are
`simpler'  than moments: the $n$-th cumulant of the conductance
distribution is represented by the sum of all the diagrams made of
$ n$ conductance loops  connected  upon the averaging over
disorder, while the $n$-th moment is the sum of both connected and
disconnected diagrams, the latter resulting in the contribution of
the lower-order cumulants in \Eq{irreduc}. For
magnetofingerprints, when the moments rather than cumulants are
naturally defined, \Eq{gBn}, one can use the inverse\cite{AdvStat}
of
 \Eq{irreduc} to express $\gBc n$ via $\gB n$:
\begin{equation*}
\frac{\gBc n}{n!} =\sum_{\{m_{i}\}}(-1)^\rho(\rho - 1)!\left\{
\prod_{k=1}^n\frac{1}{m_{k}!}\left( \frac{\gB k}{k!}\right)
^{\!m_k}\right\}, \label{reduc}
\end{equation*}%
where $\rho \equiv m_1+m_2+\ldots +m_k$.

Then a straightforward route to \Eq{result} would be to average
over disorder the square of $\gBc n$ above. However, such a route
would be  very cumbersome due to a large number of cancellations
of non-leading contributions.

Instead, we will use a so called moment-generating
function,\cite{AdvStat} defined for any well-convergent
distribution $f(g)$ by
\begin{equation}\label{mgf}
    F(t)=\int_{-\infty }^{\infty }\e^{gt}f (g)\,\dd g\,.
\end{equation}
The moments are given by coefficients of Taylor's expansion of
$F(t)$, while the cumulants are generated by
\begin{equation}\label{cum}
    \ln F(t) = \sum \gN n \frac{t^n}{n!}\;\Rightarrow\;
    \gN n=    \partial_t^n\ln F(t)\Big|_{t=0}\,.
\end{equation} We use this expression to obtain directly the
disorder-averaged variance of $\gBc n$. It is more convenient to
deal with central moments of the distribution, i.e.\ with moments
of  $\delta g(B)= g(B)- g_0(B)$, for which  the disorder-averaged
$\av{\delta g(B)}=0$. Such a shift of the distribution center does
not affect the cumulants. To simplify notations, we will still
write $\gB n $ instead of $\avB{{\delta}g^n(B)}$.

Then the variance $\text{var}\left[\gBc n\right]\equiv \av {\gBc
n^{\!\!2}} - \big\langle\gBc n  \big\rangle ^{\!2}$ of
sample-to-sample fluctuations of $\gBc n$ is given by
\begin{equation}\label{var}
\text{var}\left[\gBc n\right] = \partial _{t}^{n}\partial _{{\tau}
}^{n}\big\langle \ln F(t)\!\times\! \ln F({\tau}
)\big\rangle
 \bigg|_{\scriptstyle\begin{matrix}
  \scriptstyle{t=0} \\[-5pt]
  \scriptstyle{\tau =0} \\
\end{matrix} } \:,
\end{equation}%
where the moment-generating function (\ref{mgf}) for the
distribution (\ref{gBn}) is given by
\begin{equation}\label{F} F(t)
= \int_{{\tilde B}}^{{\tilde B}+B_{0}}\e^{g(B)t}\frac{\dd
B}{B_{0}}\,.
\end{equation}
We keep small non-Gaussian corrections only in the l.h.s.\  of
\Eq{var},   representing the disorder averaging in the r.h.s.\ as
that over the normal distribution  of $g(B)$:
\begin{gather*}
\big\langle A[g(B)]\big\rangle  \equiv  \frac1{Z}\int {\mathcal{D} g(B)}%
\,A[ g(B)]\,\e^{-{\mathcal S} }, \\
{\mathcal S}   = \frac{1}{2}\int \dd B\dd B^{\prime }
g(B)\hat{\kappa}^{-1}  g(B^{\prime })\,,\quad Z\equiv\int
{\mathcal{D} g(B)}\,\e^{-{\mathcal S} },
\end{gather*}
where   $\hat{\kappa}%
^{-1}$ is the resolvent of the correlation function  (\ref{Covar})%
. Equivalently, it can be expressed in terms of the Fourier transforms $%
g_{\omega }$ and $\kappa _{\omega }$ of a random $g(B)$ and the
correlation function $\kappa (B-B^\prime )$, with $g_{-\omega
}=g_{\omega }^{\ast }$ and $\kappa _{-\omega }=\kappa _{\omega
}^{\ast }$:
\begin{equation}\label{FT}
\left\langle A\right\rangle =\frac{1}{Z}\underset{\omega }{\prod }%
\int \!\!\dd g_{\omega }\, {A}[g_{\omega }]\,\e^{-\,\tfrac{1}{2}\int \dd\omega \,
{%
\left\vert g_{\omega }\right\vert ^{2}} {\kappa^{-1} _{\omega }}},
\end{equation}

To calculate the logarithm of the moment-generating function we
use the standard replica trick \cite{EdAn}:
\begin{equation*}
\ln F=\lim_{N\rightarrow 0}\frac{F^{N}-1}{N}\,.
\end{equation*}%
Using also \Eq{F}, we represent the variance (\ref{var}) as
\begin{gather}\label{varn}
\mathrm{var}\left[ \langle \left\langle g^{n}\right\rangle \rangle _{B}%
\right] =\lim_{N,M\rightarrow 0} \, \frac{X_{NM}^{(n)}}{NM}
\end{gather}
where
\begin{gather}
X_{NM}^{(n)} =%
\partial _{t}^{n}\partial_{\tau}^{n}
\int \!\!\frac{\mathcal{D}  g(B)}{Z}\e^{-{\mathcal S} }\!\!\int_B
\e^{t\!\sum_i g(B_{i})\,+\,{\tau} \!\sum_k g(B_{k}^{\prime })}
 \Bigg|_{\scriptstyle\begin{matrix}
  \scriptstyle{t=0} \\[-5pt]
  \scriptstyle{\tau =0} \\
\end{matrix} } \notag\:,
\\[-4pt]\label{I}\\[-10pt]\notag
\int_B\equiv \idotsint\limits_{\tilde{B}}^{\tilde{B}+B_{0}}\prod\limits_{i=1}^N%
\frac{\dd B_{i}}{B_{0}}\prod\limits_{k=1}^M\frac{\dd B_{k}^{\prime
}}{B_{0}}\,.\notag
\end{gather}
 To calculate $X_{NM}^{(n)}$ we
perform the integration over $\mathcal{D}g$ first, turning it into
the Gaussian integral over the Fourier components of $g$, \Eq{FT}:
\begin{align*}
    &\frac{1}{Z}{\prod_{\omega}  }\int \!\!\dd g_{\omega
}\,\e^{-\,\tfrac{1}{2}\int \dd\omega \,
{%
\left\vert g_{\omega }\right\vert ^{2}} {\kappa^{-1} _{\omega }}
-g_{\omega }\left( t\Sigma e^{i\omega B_{i}}+\tau\Sigma e^{i\omega
B_{k}^{\prime }}\right)}\\&= \exp \left\{ \int  \!\!\dd {\omega
}\,\left[ \frac{\kappa _{\omega }}{2}\big\vert t\Sigma e^{i\omega
B_{i}}+\tau\Sigma e^{i\omega B_{k}^{\prime }}\big\vert ^{2}\right]
\right\} .
\end{align*}
After the inverse Fourier transform for $\kappa_{\omega}$, we
arrive at
\begin{equation*}
X_{NM}^{(n)}=\partial _{t}^{n}\partial_{\tau}^{n}
\int_B\exp\left[{ t^{2}V_{BB}+{\tau} ^{2}V_{B'B'}+t{\tau} V_{BB'
}}\right] \Bigg|_{\scriptstyle\begin{matrix}
  \scriptstyle{t=0} \\[-5pt]
  \scriptstyle{\tau =0}
\end{matrix} } \:,\\[-12pt]
\end{equation*}%
where%
\begin{subequations}\label{V}
\begin{align}
&V_{BB}\mspace{-40mu} &&=\mspace{9mu}\frac{N}{2}\kappa
(0)+\sum_{i=1}^N\sum_{j=i+1}^N\kappa (B_{i}-B_{j})
\label{BB}\\
&V_{B'B'} \mspace{-40mu} &&=\mspace{9mu}\frac{M}{2}\kappa
(0)+\sum_{k=1}^{ M}\sum_{m=k+1}^M\kappa (B_{k}^{\prime
}-B_{m}^{\prime })  \label{BpBp}  \\
&V_{BB'} \mspace{-40mu} &&=\mspace{9mu}\phantom{\frac{M}{2}\kappa
(0) + \,}\sum_{i=1}^N\sum_{k=1}^{ M} \kappa (B_{i}-B_{k}^{\prime
}) \,. \label{BBp}
\end{align}
\end{subequations}
The derivatives with respect to $t$ and ${\tau} $ in the limit
$t,{\tau} \to0$ in the expression for $X_{NM}^{(n)}$ above allows
one to keep only the terms proportional to $t^n{\tau} ^m$ in the
Taylor expansion of the exponent,
 which leads to
\begin{equation}\label{X}
 X_{NM}^{(n)}= (n!)^2\int_B\,
 \sum_{l}\frac{ V_{BB'}^{n-2l}V_{BB}^{l}V_{B'B'}^{l}}{(n-2l)!(l!)^{2}}%
\,.
\end{equation}

The expression above is entirely equivalent to one that can be
obtained directly from diagrammatic techniques. The replica trick,
as always in perturbative calculations, is an exact tool that
serves to eliminate automatically spurious diagrams, in the
present case those that contribute to reducible moments but not to
cumulants.

The next step of the calculation consists in selecting
contributions of the lowest power in $B_c/B_0$  from \Eq X. Each
term in the sum contains a product of $n$ correlation functions
(\ref{Covar}), each being dependent on a difference between two
$B$'s from the multiple integral (\ref I). The integration will
produce a factor of $(B_c/B_0)^m$, where $m$ is the number of {\it
different} pairs of $B$'s in the product.

To illustrate this point, let us first consider the term with
$l=0$ in \Eq X, proportional to the $n$-th power of the sum of
$NM$ different ${\kappa}$'s  in \Eq{BBp}. The leading contribution
comes from the sum of the $n$-th power of each term in \Eq{BBp},
as each contains only one difference between different $B$'s; as
they are all equivalent,   re-labelling   reduces the $l=0$
contribution in \Eq X to
\begin{equation}\label{R}
\frac{X_{NM}^{(n0)}}{NM}\to n!\, \int_B{\kappa}^n\lr(){B-B'}=
\frac{B_c}{B_0} a_n \gN n^2\,,
\end{equation}
with $a_n$ given in \Eq{result}, since the multiple integral (\ref
I) is reduced to a single integral over $x=(B\!-\!B')/B_0$,
proportional to $B_c/B_0$. All other contributions from the $l=0$
term are proportional to $NM\,P_{N,M}$ (where $P_{N,M}$ is regular
in the replica limit), but they contain more integrals over powers
of $K$, each adding an extra small factor of $B_c/B_0$.

Now we show that the terms with $l\ne0$ in \Eq X do not contribute
in the leading order in $B_c/B_0$. Schematically, their
contributions may be represented as
\begin{align}\label{S}
    {\Sigma} _{NM}^{n-2l}\lr(){{\alpha} N+{\Sigma} _{N(N-1)}}^l
    \lr(){{\alpha} M+{\Sigma} _{M(M-1)}}^l\,,
\end{align}
where ${\Sigma} $'s with appropriate indices stand for one of the
sums in Eqs.~(\ref V). The number of independent integrations over
$\kappa$'s is not smaller than the number of different ${\Sigma}
$'s: therefore, contributions from \Eq S that contain different
${\Sigma} $'s will be small in powers of $B_c/B_0$. On the other
hand, those which contain only one ${\Sigma} $, like ${\Sigma}
_{NM}^{n-2l}({\alpha} ^2NM)^l$ or   $({\alpha} N)^{n/2}{\Sigma}
_{M(M-1)}^{n/2}$, would vanish in the replica limit in \Eq{varn},
the latter term vanishing due to the fact that each ${\Sigma} $ is
proportional to (at least) the number of the terms in it,
indicated by its index.

Therefore, \Eq R gives the only non-vanishing in the replica limit
term proportional to the first power of $B_c/B_0$. Substituting it
into \Eq{varn}, we obtain the result of \Eq{result}. Let us
discuss its quantitative implications.

First we express $K(x)$ in \Eq{Covar}  as
\cite{AlKh:85,L+S:85,Falko:92,AB}
\begin{equation}\label{K}
K(x)=\ \left[ 1+x^{2}\right] ^{-\gamma}\,,\quad x\equiv
{\Delta}B/B_c\,,
\end{equation}
for the  three types of the conductors  with different values of
the exponent ${\gamma} $ and the correlation field $B_c$:
 \begin{enumerate}[(a)]\item ${\gamma}\!=\!1/2$ and
 $B_{c}\!\sim\! \phi _{0}/[w\times \min \{L_{\varphi },L_{s}\}]$
 for a long wire
of width $w\!\ll \!\min \{L_{\varphi },L_{s}\}\!<\!L$, where
$\phi_0\equiv h e/c$
 \item
${\gamma}\!=\!1$  and $B_{c}\sim \sqrt{\frac{N\Delta
}{E_{Th}}}\phi _{0}/L^{2}$ for a dot of size $L$ connected to the
reservoirs via leads with $N$ channels and mean level spacing
$\Delta $ at temperature $T>N\Delta $
 \item ${\gamma} \!=\!
2$  and $B_{c}\sim \phi _{0}/L^{2}$ for a dot or a short wire
(with $L<\min \{L_{\varphi },L_{s}\}$) at $T=0$.
\end{enumerate}
For these  cases, coefficients $a_n$ in \Eq{result} are given by
\begin{equation*}
a_n=\left\{\begin{array}{cl}
\frac{2^{n}n}{n-1}\left[ \Gamma (\frac{n+1}{2})\right] ^{2} & \text{(a)}\;%
\\[3pt]
n\Gamma (n-\frac{1}{2})\sqrt{\pi } & \text{(b)}
\\[3pt]
\frac{n!\Gamma (2n-\frac{1}{2})\sqrt{\pi }}{\Gamma (2n)} &
\text{\parbox{4.5cm}{(c)}}
\end{array}%
\right.
\end{equation*}

The appropriate power of the variance in the r.h.s.\ of
\Eq{result} represents an inevitable systematic error in
extracting the $n^{\text{th}}$ cumulant from a single
magnetofingerprint. This can be rather large. The widest range of
magnetic fields achievable experimentally is limited to
$B_0\lesssim 20\text{T}$ by the necessity to use noiseless
superconducting magnets. For the lowest temperatures in
experiments with metallic wires, $T \sim 10\div100\text{mK}$, the
correlations field  $B_c\sim10\div 100\text{mT}$
\cite{Chang:96a,Marcus:96a,MW}, in agreement with the estimates
for type (a) above. A meaningful deviation of, say, $\gN 3$ from
the combined prediction of \EQS{gn}{Lphi} (not speaking of a more
accurate estimate of the third cumulant \cite{Macedo} that shows
it to be of order $1/g^2$ rather than $1/g$) which would manifest
a breakdown of the one-parameter scaling, or inadequacy of the
model of noninteracting electrons in disordered potential should
therefore comfortably exceed the value of $0.15\gN 2$. So far,
reliable experimental data for the higher cumulants in the
metallic regime fall well below this limit.

We thank Boris Altshuler and George Pickett for useful
discussions. This work, which was initiated during workshop
INT-02-2 'Chaos and Interactions: from Nuclei to Quantum Dots' in
the Institute for Nuclear Physics at the University of Washington.
It has further been supported by EPSRC (VIF and IVL) and Packard
Foundation (ILA). VIF also acknowledges Royal Society for travel
support.


\begin{thebibliography}{10}

\bibitem{Al:85}
B.~L. Altshuler, {\it JETP Letters} {\bf 41}, 648 (1985).

\bibitem{AlKh:85} B.~L. Altshuler and D.~E. Khmelnitskii, {\it ibid} {\bf
42}, 359
  (1985).

\bibitem{L+S:85}
P.~A. Lee and A.~D. Stone, {\it Phys. Rev. Lett.} {\bf 55}, 1622
(1985); P.~A. Lee, A.~D. Stone, and H.~Fukuyama, {\it Phys. Rev.
{\rm B}} {\bf 35},
  1039 (1987).

\bibitem{ALW} For earlier reviews of both theory and experiment, see {\it
Mesoscopic Phenomena in Solids}, {ed.\ by} B.~L. Altshuler, P.~A.
Lee,
  and R.~A. Webb, Elsevier, Amsterdam (1991).

\bibitem{Chang:96a}
A.~M. Chang et~al., {\it Phys. Rev. Lett.} {\bf 76}, 1695 (1996).

\bibitem{Marcus:96a}
J.~A. Folk et~al., {\it Phys. Rev. Lett.} {\bf 76}, 1699 (1996).

\bibitem{MW}
P.~Mohanty and R.~A. Webb, {\it Phys. Rev. Lett.} {\bf 88}, 146601
(2002).

\bibitem{Swastik}
S.~Kar, A.K.Raychaudhuri, and A.~Ghosh, cond-mat/0212165.

\bibitem{AdvStat}
M.~G. Kendall and A.~Stuart, {\it The advanced theory of
statistics,
  $4^{\text{th}}$ {\em ed., Vol.1}}, Griffin, London (1976).

\bibitem{AALR}
E.~Abrahams et al., {\it
  Phys. Rev. Lett.} {\bf 42}, 673 (1979).

\bibitem{AKL:86}
B.~L. Altshuler, V.~E. Kravtsov, and I.~V. Lerner,  Zh.\ Eksp.\
Teor.\ Fiz.\
  {\bf 91}, 2276 (1986); {in} \Ref{ALW}, p.449.

\bibitem{MPS}
K.~A. Muttalib, J.~L. Pichard, and A.~D. Stone, {\it Phys. Rev.
Lett.} {\bf
  59}, 2475 (1987).

\bibitem{comment1}
There exist contributions to higher cumulants that lead to
lognormal tails of
  the distribution function \cite{AKL:86,EF2} and are not described within the
  one-parameter scaling approach; they become larger than those in \Eq{gn} only
  for $n>\av g\gg1$ so that they are irrelevant for considerations in
  this paper.



\bibitem{comment2}
For the conductance distribution in quasi-1D wires, a
theoretically expected
  deviation from the Gaussian shape is even smaller than predicted by \Eq{gn},
  since $A_{31}=0$ so that the third cumulant turns out to
  be\cite{Macedo}
  of order $1/g^2$ rather than $1/g$.

\bibitem{comment3}
We keep only the leading terms in \Eq{Lphi}; assuming independent
contributions
  of $N$ connected in series conductors making the wire of length $L=NL\phi$
  and $\gN n\ll\av g^n$, one obtains more detailed
  expressions, e.g.\ for $n=2,3$, \[\begin{aligned} \gN
  2_{_{L=NL_\phi}}&=\frac{\gN 2}{N^3}- \frac{2(N\!-\!1)}{N^4}\,\frac{\gN 3}{\av
  g}+3\frac{(N\!-\!1)^2}{N^5}\,\frac{\gN 4}{\av
  g^2}\\&+4\frac{(N\!-\!1)}{N^4}\,\frac{{\gN 2}^2}{\av g ^2}\\  \gN
  3_{_{L=NL_\phi}}&=\frac{\gN 3}{N^5}- \frac{3(N\!-\!1)}{N^6}\,\frac{\gN 4}{\av
  g}\,. \end{aligned} \]As this is irrelevant for further considerations, we also
  disregard here the averaging over spectra which would lead to a substitution
  of the thermal length $L_T=\sqrt{D/T}$ for some of the phase breaking lengths
  $L_\phi$ in the above expansion.

\bibitem{Sanquer}
D.~Mailly and M.~Sanquer, {\it J.Phys.I (France)} {\bf 2}, 357
(1992).

\bibitem{comment4}
To exclude effects of weak localization (or antilocalization), in
  determining the conductance distribution from $g(B)$ one  disregards
  small magnetic fields, choosing $\tilde B\!\gg\! B_c$.

\bibitem{Falko:92}
V.~I. Fal'ko,  {\it Phys. Rev. {\rm B}} {\bf 51}, 5227 (1995).

\bibitem{AB}
In the 1D case, \Eq K for ${\gamma} \!=\!1/2$ is a numerically
accurate approximation to the analytic result of  I.~L. Aleiner
and Y.~M. Blanter, {\it Phys. Rev. B} {\bf 65}, 115317 (2002).

\bibitem{AKL:86a}
B.~L. Altshuler, V.~E. Kravtsov, and I.~V. Lerner, {\it JETP
Letters} {\bf 43},
  441 (1986).

\bibitem{EdAn}
S.~Edwards and P.~W. Anderson, {\it J. Phys. {\rm F}} {\bf 5}, 965
(1975).

\bibitem{EF2}
V.~I. Fal'ko and K.~B. Efetov, {\it Europhys. Lett.} {\bf 32}, 627
(1995); {\it Phys. Rev. {\rm B}} {\bf 52}, 17413 (1995).

\bibitem{Macedo}
A.~M.~S. Mac\^edo, {\it Phys. Rev. {\rm B}} {\bf 49}, 1858 (1994);
M.~C.~W. van Rossum et~al.,
  {\it Phys. Rev. {\rm B}}  {\bf 55}, 4710 (1997).

\end{thebibliography}

\end{document}